\documentclass[aps,showpacs,twocolumn]{revtex4}%
\usepackage{amsfonts,amsmath,amssymb}
\usepackage{graphicx}%
\begin{document}
\preprint{HEP/123-qed}
\title{Strong suppression of Coulomb corrections to the cross section of $e^+e^-$ pair production in ultrarelativistic nuclear collisions}
\author{R.N. Lee}
\author{A.I. Milstein}
\affiliation{The Budker Institute of Nuclear Physics and
 Novosibirsk State University,630090, Novosibirsk, Russia}


\begin{abstract}
The Coulomb corrections to the cross section of $e^+e^-$ pair production in ultrarelativistic nuclear collisions are calculated in the next-to-leading approximation with respect to the parameter $L=\ln \gamma_A\gamma_B$ ($\gamma_{A,B}$ are the Lorentz factors of colliding nuclei). We found considerable reduction of the  Coulomb corrections even for large $\gamma_A\gamma_B$ due to the suppression of the production of $e^+e^-$ pair with the total energy of the order of a few electron masses in the rest frame of one of the nuclei. Our result explains why the deviation from the Born result were not observed in the experiment at SPS \cite{Vane1994,Vane1997}.
\end{abstract}
\maketitle

Electron-positron pair production in ultrarelativistic nuclear collisions is investigated intensively during almost two last decades, see recent reviews \cite{Baur2007,Hencken2008}. This process is important in the problem of beam lifetime and luminocity of hadron colliders. It is also a serious background for many experiments because of its large cross section. For heavy nuclei, the effect of higher-order terms (Coulomb corrections) of the perturbation theory with respect to the parameters $Z_A\alpha$ and $Z_B\alpha$ can be very important ($Z_A$ and $Z_B$ are the charge numbers of the nuclei $A$ and $B$, $\alpha\approx 1/137$ is the fine-structure constant). However, no evidence of the Coulomb corrections has been found in the experiments \cite{Vane1994,Vane1997}. This circumstance stimulated considerable theoretical interest to this process. In the set of theoretical works \cite{Baltz1998,Segev1999,Eichmann1999} it was found that the exact in $Z_{A,B}\alpha$ cross section coincides with that obtained in the Born approximation in the ultrarelativistic limit. This statement was considered as an explanation of the experimental results
\cite{Vane1994,Vane1997}. However, this conclusion contradicted to the result obtained in Ref. \cite{Ivanov1999} with the help of the Weizs\"acker-Williams approximation in the leading logarithmic approximation. This contradiction has been resolved in Ref. \cite{Lee2000a}. It was shown that the wrong conclusion of Refs. \cite{Baltz1998,Segev1999,Eichmann1999} on the absence of Coulomb corrections was due to the bad treatment of conditionally convergent integrals. Consistent approach of Ref. \cite{Lee2000a} results in the Coulomb corrections which coincide with those from Ref. \cite{Ivanov1999}. Thus absence of the Coulomb corrections in the experiments \cite{Vane1994,Vane1997} has remained unexplained.
In the present paper, we explain this experimental result by the suppression of the Coulomb corrections due to the account of the next-to-leading term.

Since the nuclear mass is large compared to the electron mass, it is possible to treat the nuclei as sources of the external field and calculate the probability $P_n(b)$ of $n$-pair production at a fixed impact parameter $b$. It is convenient to introduce the average number $W(b)$ of produced pairs and the number-weighted cross section $\sigma_T$ as

\begin{equation}
    W(b)=\sum_{n=1}^{\infty}nP_n(b)\,,\quad
    \sigma_T=\int d^2b W(b)=\sum_{n=1}^{\infty}n\sigma_n\,,
\end{equation}
where $\sigma_n=\int d^2b P_n(b)$ is the cross section of $n$-pair production. The cross section $\sigma_T$ can be presented in the form:
\begin{equation}  \label{sigmaccc}
\sigma_T=\sigma^0+\sigma^A+\sigma^B+\sigma^{AB} \, ,
\end{equation}
where $\sigma^0\propto(Z_A\alpha)^2(Z_B\alpha)^2$ is the Born cross section, $\sigma^A$ and $\sigma^B$ are the Coulomb corrections with respect to nucleus $A$ and $B$, respectively (containing the terms proportional to $(Z_B\alpha)^{2}(Z_A\alpha)^{2n}$ and $(Z_B\alpha)^{2n}(Z_A\alpha)^{2}$, $n\geqslant 2$), and $\sigma^{AB}$ is the Coulomb corrections with respect to both nuclei (containing the terms proportional
to $(Z_B\alpha)^{n}(Z_A\alpha)^{l}$ with $n,l> 2$). The cross section $\sigma^0$ coincides with the Born  cross section of one pair
production, which was calculated many years ago in Refs.
\cite{Landau1934,Racah1937}. In the leading logarithmic approximation, the quantities $\sigma^{A,B}\propto L^2$ and $\sigma^{AB}\propto L$  were obtained in Refs. \cite{Ivanov1999,Lee2000a} and Ref. \cite{Lee2001b}, respectively.

The leading logarithmic approximation for $W(b)$ provides the factorization of $P_n(b)$
\cite{Baur1990,Rhoades-Brown1991,Best1992,Hencken1995}, so that
\begin{equation}\label{Pn}
P_n( b)=\frac{W^n( b)}{n!}\mathrm{e}^{-W( b)}\,.
\end{equation}
The function $W(b)$ was calculated in Refs. \cite{Hencken1995a,Gucclu1995,Gucclu2000,Lee2002,Hencken2004} in the Born approximation and  in Refs. \cite{Alscher1997,Hencken1999,Baltz2006,Lee2007a} with the Coulomb corrections taken into account. Using Eq. (\ref{Pn}), the cross section $\sigma_1$ of one pair production can be represented as a sum of $\sigma_T$ and the unitarity correction $\sigma_{\mathrm{unit}}$
\begin{eqnarray}\label{sigma1W}
\sigma_1&=&\sigma_T+\sigma_{\mathrm{unit}}\,,\nonumber\\
\sigma_{\mathrm{unit}}&=& -\int d^2{\ b}W( b) \left(1-\mbox{e}^{-W( b)}\right) \, .
\end{eqnarray}
The existence of the unitarity correction was first recognized in Ref. \cite{Baur1990a} (see also review \cite{Baur2007}). Numerical evaluation of this correction was performed in Refs. \cite{Lee2002,Jentschura2008}.
The main contribution to $\sigma_1$ is given by the term $\sigma^0$ in $\sigma_T$, Eq. (\ref{sigmaccc}), and is known with high accuracy \cite{Landau1934,Racah1937}. The terms $\sigma^A$ and $\sigma^B$ in $\sigma_T$ also give important contributions to $\sigma_1$. In the leading logarithmic approximation, these terms have been derived in Refs. \cite{Ivanov1999,Lee2000a}. The last two contributions, $\sigma^{AB}$ and $\sigma_{\mathrm{unit}}$, to $\sigma_1$ are rather small, see Refs. \cite{Lee2001b,Lee2002}.

In the present paper, we calculate the leading corrections to $\sigma^{A,B}$ (which are also the corrections to $\sigma_1$). We show that these corrections essentially diminish the magnitude of $\sigma^{A,B}$ even for the parameters of LHC ($\gamma_A=\gamma_B\approx 3000$, $Z_A=Z_B=82$). It is convenient to calculate $\sigma^A$ in the rest frame of the nucleus $A$, where the nucleus $B$ has the Lorenz factor $\gamma = 2\gamma_A\gamma_B$ at $\gamma_{A,B}\gg 1$. Note that $\sigma^A$, being proportional to $(Z_B\alpha)^2$, can be directly calculated as the Coulomb corrections to $\sigma_1$ with respect to the parameter $Z_A\alpha$, so that it can be represented as
\begin{eqnarray}\label{eq:sig1}
\sigma^A&=&\int_{2m}^{\infty} d\omega \int_{(\omega/\gamma)^2}^{\infty} dQ^2\left[
   \frac{dn_{\perp}(\omega,Q^2)}{d\omega dQ^2}\sigma_{\perp}(\omega,Q^2)\right.\nonumber\\
   &&\left. +\frac{dn_{\parallel}(\omega,Q^2)}{d\omega{}dQ^2}\sigma_{\parallel}(\omega,Q^2)
\right]
\end{eqnarray}
Here
\begin{eqnarray}\label{eq:dn1}
    dn_{\perp}(\omega,Q^2)&=&\frac{Z_B^2\alpha}{\pi}\left(1- \frac{(\omega/\gamma)^2}{Q^2}\right) \frac{d\omega}{\omega}\frac{dQ^2}{Q^2}\,,
\nonumber\\
    dn_{\parallel}(\omega,Q^2)&=&\frac{Z_B^2\alpha}{\pi} \frac{d\omega}{\omega}\frac{dQ^2}{Q^2}.
\end{eqnarray}
are the numbers of virtual photons  $\gamma^*_{\perp,\parallel}$ with the energy $\omega$, the  virtuality $-Q^2<0$, and the transverse and longitudinal polarizations, respectively. The quantities $\sigma_{\perp}(\omega,Q^2)$ and $\sigma_{\parallel}(\omega,Q^2)$ are the Coulomb corrections to  the cross sections of the processes $\gamma^*_{\perp,\parallel} A\to e^+e^-A$.

Let us discuss the contributions to $\sigma^A$ of different regions of integration with respect to $\omega$ and $Q^2$.

The leading logarithmic contribution $\propto L^2$ comes from the integration of $\sigma_{\perp}$ over the region
\begin{equation}
    \mathrm{I.}\quad m\ll \omega\ll m\gamma\,,\quad (\omega/\gamma)^2\ll Q^2 \ll  m^2\,.\label{eq:Region1}
\end{equation}
The leading correction  $\propto L$ comes from  the following regions:
\begin{eqnarray}
\mathrm{II.}&\quad&Q^{2}    \sim m^{2}\,,\quad m\ll\omega\ll\gamma m\\
\mathrm{III.}&\quad&Q^{2}    \sim\left(  \omega/\gamma\right)  ^{2}%
\,,\quad m\ll\omega\ll\gamma m\\
\mathrm{IV}.&\quad&\omega   \sim m\,,\quad\left(  m/\gamma\right)  ^{2}\ll Q^{2}\ll
m^{2}%
\end{eqnarray}
Note that the cross section $\sigma_{\parallel}$ gives logarithmically enhanced contribution only in region  $\mathrm{II}$. Therefore, since we are going to keep the terms proportional to $L^2$ or $L$, we can write $\sigma^A$ as
\begin{eqnarray}
&&\sigma^A=\sigma_{\mathrm{as}}^A+\delta\sigma^A\,,\nonumber\\
&&
\label{eq:sig2}
\sigma_{\mathrm{as}}^A=
\int_{2m}^{\infty} d\omega \int_{(\omega/\gamma)^2}^{\infty} dQ^2\left[\frac{dn_{\perp}(\omega,Q^2)}{d\omega dQ^2}
   \sigma_{\perp}(\infty,Q^2)\right.\nonumber\\
&&
\left.+\frac{dn_{\parallel}(\omega,Q^2)}{d\omega{}dQ^2}\sigma_{\parallel}(\infty,Q^2)
\right]\,,\\
&&
\label{eq:sig3}
\delta\sigma^A=
\int\limits_{2m}^{\infty} d\omega \int\limits_{(\omega/\gamma)^2}^{\infty} dQ^2\frac{dn_{\perp}(\omega,Q^2)}{d\omega dQ^2}\delta\sigma_{\perp}(\omega,Q^2)\nonumber\\
&&
\delta\sigma_{\perp}(\omega,Q^2)=
\sigma_{\perp}(\omega,Q^2)-\sigma_{\perp}(\infty,Q^2)\,.
\end{eqnarray}

The quantities $\sigma_{\perp,\parallel}(\infty,Q^2)$ can be calculated within the quasiclassical approximation. Following the method described in detail in Ref. \cite{Lee2004}, we obtain
\begin{gather}
\sigma_{\perp,\parallel}(\infty,Q^2)=\frac{\alpha}{\omega} \mathrm{\operatorname{Re}}\int d\varepsilon\, \mathrm{\operatorname{Sp}}\int d\mathbf{r}_{1}d\mathbf{r}_{2}e^{-i\mathbf{k} \cdot\mathbf{r}}\nonumber\\
\times
\left[  \left(  2\mathbf{e}\cdot\mathbf{p}_{2} +\hat{k}\hat{e}\right)  D_{-}\right]
\left[  \left(  2\mathbf{e}^{\ast}\cdot\mathbf{p}_{1}-\hat{k}\hat{e}^{\ast}\right)  D_{+}\right]\,,\nonumber\\
D_{-}=D(\mathbf{r}_{2},\mathbf{r}_{1}| \varepsilon)\,,\quad
D_{+}=D(\mathbf{r}_{1},\mathbf{r}_{2}| \varepsilon-\omega)\,,
\end{gather}
where $D(\mathbf{r}_{2},\mathbf{r}_{1}| \varepsilon)$ is the quasiclassical Green function of the squared Dirac equation, $e=(0,1,0,0)$ for $\sigma_\perp$ and $e=(0,0,0,Q/\omega)$ for $\sigma_\parallel$ in the frame where $\mathbf{k}$ is directed along $z$ axis. Using the explicit expressions for the Green functions from Ref. \cite{Lee2004}, we obtain the results for these cross sections:
\begin{eqnarray}\label{eq:sig4}
\sigma_\perp(\infty,Q^2)&=&\alpha N\int\limits_0^1 dy\frac{1+2\left(1-2y\bar{y}\right)
(1+y\bar{y}Q^2/m^2)}{(1+y\bar{y}Q^2/m^2)^2}\,,\nonumber\\
\sigma_\parallel(\infty,Q^2)&=&4\alpha N\int\limits_0^1 dy
\frac{y^2\bar{y}^2Q^2/m^2}{(1+y\bar{y}Q^2/m^2)^2}\,,\nonumber\\
N&=&-\frac{4(Z_A\alpha)^{2}}{3m^2}
\mathrm{\operatorname{Re}}
[\psi(1+iZ_A\alpha)-\psi(1)]\,,\nonumber\\
\bar{y}&=&1-y\,,
\end{eqnarray}
where $\psi(x)=d\ln\Gamma(x)/dx$. These formulas agree with the result of Ref. \cite{Ivanov1998a} if one takes into account the missing factor $y\bar{y}$ in $\sigma_\parallel$ pointed out in Ref. \cite{Ivanov1998}. Substituting Eq. (\ref{eq:sig4}) in Eq.  (\ref{eq:sig2}) and taking the integrals over $\omega$ and $Q^2$, we obtain within the logarithmic accuracy
\begin{eqnarray}\label{eq:sig_as}
    \sigma_{\mathrm{as}}^A&=&
\frac{7(Z_B\alpha)^2N}{3\pi}
\left[ L^2+\frac{20}{21}L \right]\,.
\end{eqnarray}
We remind that $L=\ln(\gamma_A\gamma_B)=\ln(\gamma/2)$. The result (\ref{eq:sig_as}) is in agreement with those obtained in Refs. \cite{Ivanov1998,Gevorkyan2003}.

Let us pass to the contribution $\delta \sigma^A$, Eq. (\ref{eq:sig3}), which was not considered so far. In Ref. \cite{Gevorkyan2003} it was conjectured that the term $\delta\sigma^A$ can be safely omitted. We show below that this guess is completely wrong. The function $\delta\sigma_{\perp}(\omega,Q^2)$ in the integrand provides convergence of the integral over $\omega$ in the region $\omega\sim m$. The logarithmically enhanced contribution is given by the region
$(m/\gamma)^{2}\ll Q^{2}\ll m^{2}$ of integration over $Q^2$. Since $Q^2\ll m^2$ in this region, we can make the substitution $\delta\sigma_{\perp}(\omega,Q^2)\to\delta\sigma_{\perp}(\omega,0)$ in Eq. (\ref{eq:sig3}). Then we take the integral over $Q^2$ and obtain
\begin{eqnarray}\label{eq:delta_sig}
\delta\sigma^A&=&
\frac{7(Z_B\alpha)^2N\, G(Z_A\alpha)}{3\pi} L\,,\nonumber\\
G(Z_A\alpha)&=&
2\int_{2m}^{\infty}\frac{d\omega}{\omega}
\left[\frac{\sigma_{\perp}\left(\omega,0\right)}{\sigma_{\perp}\left(\infty,0\right)} -1\right]\,.
\end{eqnarray}
The quantity $\sigma_{\perp}(\omega,0)\equiv \sigma_{\gamma A}(\omega)$ is the Coulomb corrections to the cross section of $e^+e^-$ pair production by real photon in the Coulomb field, and $\sigma_{\perp}\left(\infty,0\right)=7\alpha N/3$. Taking the sum of Eqs. (\ref{eq:sig_as}) and (\ref{eq:delta_sig}), we finally obtain $\sigma^A$ in the next-to-leading approximation
\begin{eqnarray}\label{eq:sig_total}
\sigma^A&=&
\frac{7(Z_B\alpha)^2N}{3\pi}\left[ L^2+\left(G(Z_A\alpha)+\frac{20}{21}\right)L \right]\,.
\end{eqnarray}
In order to calculate the function $G(Z_A\alpha)$ it is necessary to know the magnitude of the Coulomb corrections $ \sigma_{\gamma A}(\omega)$ in the energy region where the produced $e^+e^-$ pair is not ultrarelativistic. The formal expression for it, exact in
$Z_A\alpha$ and $\omega$, was derived in Ref. \cite{Overbo1968}. This expression has a very complicated form causing severe difficulties in computations. The difficulties grow as $\omega$ increases, so that numerical results in Refs. \cite{Overbo1968,Overbo1973} were obtained only for $\omega<5 MeV$. In a set of later publications \cite{WrigSudKos1987,SudVarga1991,SudVarga1994,SudSharm2006} (see also reviews \cite{HubbeSeltz2004,Hubbell2006}) the magnitude of $\sigma_{\gamma A}(\omega)$ has been obtained for higher values of $\omega$ and several $Z_A$. In the high-energy region $\omega\gg m$, the consideration is greatly simplified. As a result, a rather simple form of the Coulomb corrections was obtained in \cite{BetheMaxim1954,DaviBetMax1954} in the leading approximation with respect to $m/\omega$ and in Ref.\cite{Lee2004} in the next-to-leading approximation. In Ref. \cite{Overbo1977}, a simple formula, which correctly reproduces the low-energy results and the high-energy limit, was suggested. This "bridging" expression has high accuracy at intermediate energies and differs from the exact result for $\sigma_{\gamma A}(\omega)$ only in the region close to the threshold $\omega=2m$. For our purpose, this difference is not important because in this region the ratio ${\sigma_{\gamma A}(\omega)}/{\sigma_{\gamma A}(\infty)}$  can be neglected in comparison with unity.

\begin{figure}
\centering\setlength{\unitlength}{0.1cm}
\begin{picture}(70,50)
 \put(37,-1){\makebox(0,0)[t]{$Z$}}
 \put(-5,25){\rotatebox[origin=c]{90}{$G(Z\alpha)$}}
\put(0,0){\includegraphics[width=70\unitlength,keepaspectratio=true]{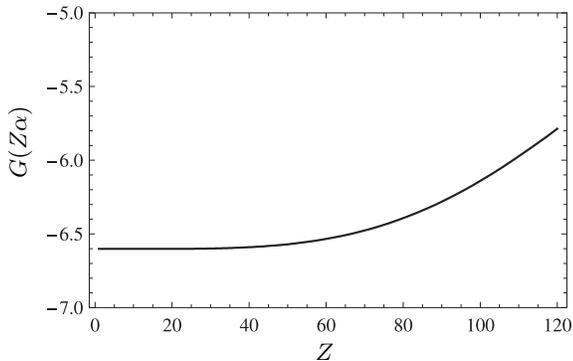}}
\end{picture}
\caption{The dependence of $G(Z\alpha)$ on $Z$.}
\label{fig:G}
\end{figure}
\begin{figure}[h]
\centering\setlength{\unitlength}{0.1cm}
\begin{picture}(70,50)
 \put(37,-1){\makebox(0,0)[t]{$\gamma$}}
 \put(-5,25){\rotatebox[origin=c]{90}{$\sigma^A/\sigma^A_{\mathrm{LA}}$}}
\put(0,0){\includegraphics[width=70\unitlength,keepaspectratio=true]{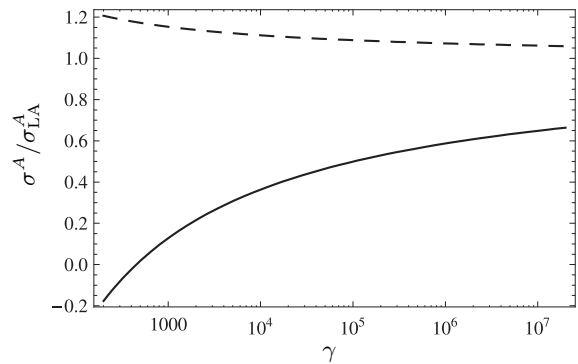}}
\end{picture}
\caption{The ratio $\sigma^A/\sigma^A_{\mathrm{LA}}$ (solid curve) as a function of $\gamma$ for $Z_A=82$. Here $\sigma^A_{\mathrm{LA}}=7(Z_B\alpha)^2N L^2/(3\pi)$ is the Coulomb corrections calculated in the leading logarithmic approximation. Dashed curve shows the ratio $\sigma^A_{\mathrm{as}}/\sigma^A_{\mathrm{LA}}$.}
\label{fig:sig_A}
\end{figure}

The function $G(Z\alpha)$ is shown in Fig. \ref{fig:G}. It is seen that $G(Z\alpha)$
varies slowly from $-6.6$ for $Z=1$ to $-6.14$ for $Z=100$ being large for all interesting values of $Z$. The large value of $G$ leads to a big difference between $\sigma^A$ from Eq. (\ref{eq:sig_total}) and its leading logarithmic approximation $\sigma^A_{\mathrm{LA}}=7(Z_B\alpha)^2N L^2/(3\pi)$ even for very large $\gamma$. This statement is illustrated in Fig. \ref{fig:sig_A} where the ratio $\sigma^A/\sigma^A_{\mathrm{LA}}$ is shown as a function of $\gamma$ (solid curve). If one omits the contribution $\delta\sigma^A$ and use $\sigma^A_{\mathrm{as}}$, Eq. (\ref{eq:sig_as}), as an approximation to $\sigma^A$, then the contribution of linear in $L$ term becomes much less important, see the dashed curve in Fig. \ref{fig:sig_A}.
Note that for $\mathrm{Pb}$-$\mathrm{Pb}$ collisions at LHC one has $\gamma\approx 1.8\times 10^7$ and  $\sigma^A/\sigma^A_{\mathrm{LA}}\approx 0.66$. For$\mathrm{Au}$-$\mathrm{Au}$ collisions at RHIC one has $\gamma\approx 2.3\times 10^4$ and  $\sigma^A/\sigma^A_{\mathrm{LA}}\approx 0.42$. For the experiments at SPS  \cite{Vane1994,Vane1997}, the Lorentz factor was $\gamma\approx 200$. Naturally, we can not use the result (\ref{eq:sig_total}) obtained in the logarithmic approximation in the region $\gamma\lesssim 500$ where the logarithmic correction to $\sigma^A$ becomes larger than the leading term $\sigma^A_{\mathrm{LA}}$. However, we can claim that, due to the strong compensation between the leading term and the correction, the Coulomb corrections $\sigma^A$ are much smaller than $\sigma^A_{\mathrm{LA}}$ at $\gamma\lesssim 500$. Therefore, this naturally explains why there was no evidence of the Coulomb corrections in the experiments \cite{Vane1994,Vane1997}.

\begin{figure}[h]
\centering\setlength{\unitlength}{0.1cm}
\begin{picture}(70,50)
 \put(37,-1){\makebox(0,0)[t]{$\gamma$}}
 \put(-5,25){\rotatebox[origin=c]{90}{$\sigma^A/\sigma^0$}}
\put(0,0){\includegraphics[width=70\unitlength,keepaspectratio=true]{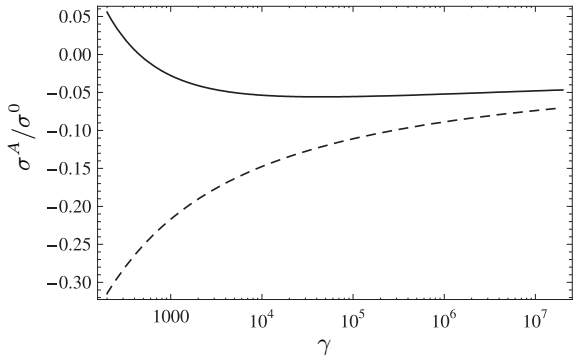}}
\end{picture}
\caption{The ratio $\sigma^A/\sigma^0$ (solid curve) as a function of $\gamma$ for $Z_A=82$. Dashed curve shows the ratio $\sigma^A_{\mathrm{LA}}/\sigma^0$.}
\label{fig:CC2Born}
\end{figure}

Let us discuss now the importance of the Coulomb corrections $\sigma^A$ in comparison with the Born cross section $\sigma^0$. The ratio $\sigma^A/\sigma^0$ is shown in Fig. \ref{fig:CC2Born}. In the next-to-leading approximation for $\sigma^A$ this ratio (solid curve) is small ($\lesssim 5\%$), while the same ratio obtained with $\sigma^A$ approximated by $\sigma^A_{\mathrm{LA}}$ reaches $20\%$ at $\gamma\sim 1000$.

To summarize, we have calculated the Coulomb corrections $\sigma^A$  to $e^+e^-$ pair production in the next-to-leading logarithmic approximation. After the account of the next-to-leading term, the magnitude of $\sigma^A$ becomes small in comparison with the Born cross section, in contrast to the leading term $\sigma^A_{\mathrm{LA}}$. The big difference between our result and previously suggested one has a simple explanation. The latter was based on the use of the high-energy asymptotics for the Coulomb corrections to  the photoproduction cross section instead of the exact Coulomb corrections. However, the exact Coulomb corrections are strongly suppressed in a rather wide region $2m<\omega\lesssim 20m$.
Note that our results, combined with $\sigma^{AB}$, Ref. \cite{Lee2001b}, complete the calculation of linear in $L$ terms in the number-weighted cross section $\sigma_T$.

This work was supported in part by RFBR Grants 09-02-00024 and 08-02-91969.

\end{document}